# Coexistence of Reconstructed and Unreconstructed Structures in Structural Transition Regime of Twisted Bilayer Graphene

Xiao-Feng Zhou[1], Yi-Wen Liu[1], Chen-Yue Hao[1], Chao Yan[1], Qi Zheng[1], Ya-Ning Ren[1], Ya-Xin Zhao[1], Kenji Watanabe[2], Takashi Taniguchi[3], and Lin He[1,*]

[1]Center for Advanced Quantum Studies, Department of Physics, Beijing Normal University, Beijing, 100875, People's Republic of China.

[2]Research Center for Functional Materials, National Institute for Materials Science, Tsukuba, Japan.

[3]International Center for Materials Nanoarchitectonics, National Institute for Materials Science, Tsukuba, Japan.

[*]Correspondence and requests for materials should be addressed to L.H. (e-mail: helin@bnu.edu.cn).

**In twisted bilayer graphene (TBG), a twist-angle-dependent competition between interlayer stacking energy and intralayer elastic energy results in flat rigid layers at large twist angles and lattice reconstruction at small twist angles. Despite enormous scientific interest and effort in the TBG, however, an experimental study of evolution from the rigid lattice to the reconstructed lattice as a function of twist angle is still missing. Here we present a scanning tunneling microscopy and spectroscopy study to reveal the twist-angle-dependent lattice reconstruction in the TBG. Our experiment demonstrates that there is a transition regime between the rigid regime and the relaxed regime and, unexpectedly, the reconstructed and unreconstructed structures coexist in the transition regime. The coexistence of the two distinct structures in this regime may arise from subtle balance between the interlayer stacking energy and intralayer elastic energy in the TBG with intermediate moiré sizes.**

In two-dimensional van der Waals (vdW) heterojunctions and homojunctions, the competition between the interlayer vdW coupling and the intralayer elastic deformation determines their stacking configurations [1-19], which have profound influences on their electronic and optical properties. For the vdW twist bilayer with a large twist angle (labelled as rigid regime), the size of the moiré pattern is quite small and two flat layers with a rigid lattice are energetically preferable. For the small twist angle case (labelled as relaxed regime), the size of the moiré pattern becomes quite large. Then, to minimize the total energy of the system, there is a strain-accompanied lattice reconstruction that results in large triangular stacking domains and a triangular network of domain walls (*DWs*) [1-12,20]. The presence of rigid lattice in large twist angle and structural reconstruction in small twist angle have been observed explicitly in a variety of vdW twist bilayers, but, despite enormous scientific interest in the vdW systems, the regime of intermediate twist angles has been the subject of substantially less experimental attention so far. Very recently, Raman measurements on a series of twist $MoS_2$ bilayers with different twist angles indicate that there is a transition regime between the rigid regime and the relaxed regime [3]. Unlike the rigid and relaxed regime, the phonon modes of the twist $MoS_2$ bilayers evolve rapidly with twist angle in the transition regime [3], suggesting that the structures of the vdW twist bilayers in the transition region are largely unexplored and out of expectation.

In this Letter, we reveal the existence of a structural transition regime, in which the reconstructed and unreconstructed structures coexist, in twisted bilayer graphene (TBG). A series of TBG with controlled twist angles $0.10° < \theta < 3.59°$ are fabricated on hexagonal boron nitride (hBN) and tungsten diselenide ($WSe_2$). By using scanning tunneling microscope (STM) and spectroscopy (STS), we directly and systematically study the structures of the TBG. Our experiment demonstrates that there is only rigid lattice for the TBG with twist angle above about $1.80°$ and only reconstructed structure for the TBG with twist angle below about $0.90°$, whereas, the reconstructed and unreconstructed structures coexist in the transition regime for the TBG with $0.90° < \theta < 1.80°$. The coexistence of the two structures in the transition regime may arise from subtle balance between the interlayer stacking energy and intralayer elastic energy in the TBG with intermediate moiré sizes.

Figure 1(a) shows schematic of the experimental device set-up. The TBG samples are obtained by both dry and wet transfer technology of graphene layer by layer on mechanical-exfoliated hBN and $WSe_2$ sheets [4,5,21–23] (using the transfer platform from Shanghai Onway Technology Co., Ltd., see Methods of the Supplemental Material for details of the device fabrication [24]). Figures 1(b) and 1(c) show schematics of atomic registries in the unreconstructed and reconstructed TBG, respectively. For the unreconstructed TBG, the stacking structure is formed by introducing a twist between two rigid graphene layers without any lattice rearrangement. For the reconstructed TBG, to reduce the total energy of the system, the *AB* and *BA* stacking regions are enlarged with separating narrow *DWs* and, simultaneously, the *AA* stacking regions are reduced. Therefore, we can distinct the unreconstructed and reconstructed TBG in two different ways in the STM measurements. The first method is based on the emergence of the *DWs* between the *AB* and *BA* stacking regions in the reconstructed TBG. The second method is based on the relative size of the *AA* and the *AB* (*BA*) stacking regions. Figures 1(d)-1(g) show several representative STM images of the obtained TBG with different twist angles in our experiment (see more STM images of the TBG in Figs. S1 and S2 [24]). The moiré superlattice can be clearly identified from the periodic corrugations in the images and the twist angles can be obtained based on the measured periods $D$ according to $D = a/[2\sin(\theta/2)]$, where $a \approx 0.246$ nm. In the 0.50° TBG, the area of the AA regions is much smaller than that of the *AB/BA* regions and there are one-dimensional (1D) *DWs* between two adjacent *AB* and *BA* regions (Fig. 1(d)), as observed previously in the reconstructed TBG with small twist angles [2,4-6,10-12]. In contrast, the structures of the 3.59° TBG can be interpreted based on a rigid lattice picture (Fig. 1(g)). Such a result is consistent with previous understanding that the TBGs with small twist angles exhibit structural reconstruction and the TBGs with large twist angles have rigid lattice structure. However, in our experiment, it is quite surprising to observe two distinct structures of the 1.09° TBG: one is in the rigid structure (Fig. 1(f)), the other is in the relaxed structure (Fig. 1(e)). This result indicates that the structure in the TBG with the intermediate twist angles (labelled as transition regime) is quite different from that of the rigid regime and the relaxed regime.

The distinct features of the two structures of the 1.09° TBG can be seen more clearly in the zoom-in STM images, as shown in Figs. 2(a) and 2(c). To quantitatively show the differences between the two structures, Figs. 2(b) and 2(d) show height profiles along dashed lines in Figs. 2(a) and 2(c) respectively, which reflect the topographic changes that pass through the *AA*, *AB* (*BA*), *DW*, *BA* (*AB*), and *AA* regions of the TBG successively. In the reconstructed TBG, a well-defined corrugated *DW* can be detected, whereas it becomes almost undetectable in the unreconstructed TBG. To systematically study the twist-angle-dependent structures in the TBG, we measure averaged $H_{DW}/H_{AA}$ as a function of the twist angles for more than 60 TBGs (here, $H_{DW}$ and $H_{AA}$ are the height of the *DW* and *AA* regions measured according to the height profiles), as summarized in Fig. 2(e). The error bars reflect the height fluctuations measured on different *AA* and *DW* regions and we should point out that the obtained result in this work is independent of the substrates. It is easy to find that the measured results can be divided into two categories: the large (small) values of the ratio $H_{DW}/H_{AA}$ indicate that the TBG has a relaxed (rigid) lattice. According to the result in Fig. 2(e), there is only reconstructed structure for the TBG with twist angles below about 0.90° and only rigid lattice for the TBG with twist angles above about 1.80°. The onset of reconstruction begins below about 1.80°, which is in good agreement with that reported in previous studies [10,20]. Unexpectedly, there is a transition regime, 0.90° < $\theta$ < 1.80°, in which both the reconstructed and unreconstructed structures coexist. Here we should point out that both the reconstructed and unreconstructed structures are quite stable during our STM measurements. The two distinct structures will result in quite different electronic properties of the TBG in the transition regime, which may provide a consistent understanding of different STS spectra observed previously in magic-angle TBG (the magic angle ~1.08° is in the transition regime) [25-29].

Besides the DWs, we also can distinguish the reconstructed and unreconstructed structures of the TBG according to the relative size of the *AA* and the *AB* (*BA*) stacking regions. However, it is quite difficult to exactly define the *AA* stacking region based on the height profiles of the STM images (see Figs. 2(b) and 2(d) as examples). To overcome this challenge, we carry out measurement of spatial distribution of the low-energy van Hove singularities (VHSs) to measure the area of the *AA* stacking region. In the TBG, a finite interlayer coupling between Dirac cones of the two adjacent layers generates a pair of saddle points, which introduce two pronounced VHSs in density of states (DOS) [30-36]. The two low-energy VHSs are mainly localized in the *AA* stacking region, especially for the TBG with twist angles $\theta < 2.0°$ [25,32-34]. Figure 3(a) shows representative STS spectra measured along the dashed line in Fig. 2(c) (here the back-gate voltage is 14 V. See STS spectroscopic map as a function of $V_{gate}$ taken at the center of an *AA* stacking region in Fig. S3 [24]). The two pronounced peaks in the spectra recorded in the *AA* stacking regions are attributed to the two low-energy VHSs of the TBG. Figure 3(b) shows a typical STS map recorded at the energy of one of the VHS, which directly reflects the spatial distribution of the DOS at the selected energy. Obviously, the electronic states of the VHS are mainly localized in the *AA* region of the TBG. Figure 3(c) displays a profile line of the DOS at the energy of the VHS along the dashed line in Fig. 3(b), which is used to estimate the size of the *AA* stacking region. To quantitatively describe the size of the *AA* stacking region, a Gaussian curve fitting is used to obtain the standard deviation σ of the profile line [37–41], as shown in Fig. 3(c) (see Supplemental Materials for details of analysis [24]). For each TBG, the σ is averaged by fitting the DOS along the three moiré directions. Figure 3(d) summarizes the averaged σ as a function of twist angles based on part of the TBG devices in Fig. 2(e). Obviously, the twist-angle-dependent sizes of the *AA* stacking regions for the reconstructed and unreconstructed TBG are quite different (the two structures of the TBG are categorized based on the results of Fig. 2). As expected, the sizes of the *AA* stacking regions in the unreconstructed TBGs increase quickly with the decrease of twist angle, whereas, the sizes of the *AA* stacking regions in the reconstructed TBGs depend weakly on the twist angle. Such a result is quite reasonable because that the area of the *AA* stacking regions in the reconstructed TBG should be significantly reduced to lower down the stacking energy of the system. Based on the result in Fig. 3(d), we also can divide the TBG with various twist angles into three regimes: the reconstructed regime for $\theta < 0.90°$, the transition regime for $0.90° < \theta <$

1.80°, and the rigid regime for $\theta > 1.80°$. Obviously, such a result is well consistent with that obtained in Fig. 2.

To further understand the observed phenomena, we calculate the stacking energies of both the reconstructed and unreconstructed TBG as a function of twist angles, according to the theoretical results in Refs. [10,42]. In the calculation, the total atoms of different TBGs are assumed to be the same and are equal to the number of atoms in a moiré unit cell, as schematically shown in Fig. 4(a), of the calculated TBG with the minimum twist angle. For the reconstructed TBG, the areas for the *AA*, *AB*, and *DW* regions in different twist angles are extracted from the results obtained in Ref. [10] (See Supplemental Materials for details of calculation [24]). For the unreconstructed TBG, two different interlayer distances are considered in the calculation and the real interlayer distance $d$ of the TBG is expected to be within them, *i.e.*, 3.2 Å $\leq d \leq$ 3.6 Å [43]. Figure 4(b) shows the calculated vdW stacking energies as a function of twist angles for both the reconstructed and unreconstructed TBG. The energy difference between the two structures is shown in Fig. 4(c) (here we use the result of $d = 3.6$ Å for the unreconstructed TBG for simplicity and the result of other distances is similar). For very small twist angles, the vdW stacking energy of the reconstructed TBGs is much smaller than that of the unreconstructed TBGs. Then, the TBG prefers to expand the *AB* and *BA* regions to gain vdW energy, which is much larger than the losing elastic energy in the formation of the *DWs*. Therefore, we always observe the reconstructed structures for the TBG with very small twist angles. With increasing the twist angle, the energy of the unreconstructed TBGs decreases slowly, while the energy of the reconstructed TBGs increases quickly. At a sufficient large twist angle, the gaining vdW stacking energy cannot compensate for the losing elastic energy in the TBG. Then, the TBG prefers to be in the structure consisting of two rigid graphene layers. As a consequence, we always observe the unreconstructed structures for the TBG with large twist angles. For the TBG with intermediate twist angles between that of the rigid regime and the relaxed regime, the gaining vdW energy and the losing elastic energy in the reconstructed TBG may be comparable. Then, the subtle balance between them makes both the reconstructed and unreconstructed TBG energetically stable. Consequently, we observe both the structures in the transition region in our experiment. In Fig. 4(c), the derivative of the vdW energy difference between the reconstructed and unreconstructed TBGs as a function of the twist angle is also plotted. It is interesting to find that the absolute value of the derivation increases rapidly below a "critical" twist angle, ~ 0.9°, which means that the energy difference between the reconstructed and unreconsturcted TBG increases much quicker below the "critical" twist angle. Such a result helps us to further understand the experimental result that there is only reconstructed TBG for $\theta <$ 0.90°.

In summary, we systematically study structures of the TBGs and demonstrate that there are three regimes categorized by their structures. The TBG is in rigid lattice for twist angles above about 1.80° and in reconstructed structure for twist angles below about 0.90°. In the transition regime, 0.90° < $\theta$ < 1.80°, both the reconstructed and unreconstructed structures coexist in the studied TBG. Our analysis indicates that subtle balance between the interlayer stacking energy and intralayer elastic energy in the TBG with intermediate moiré sizes leads to the coexistence of the two structures in the transition regime.

**Acknowledgments:**

This work was supported by the National Key R and D Program of China (Grant Nos. 2021YFA1401900, 2021YFA1400100) and National Natural Science Foundation of China (Grant Nos. 12141401,11974050).

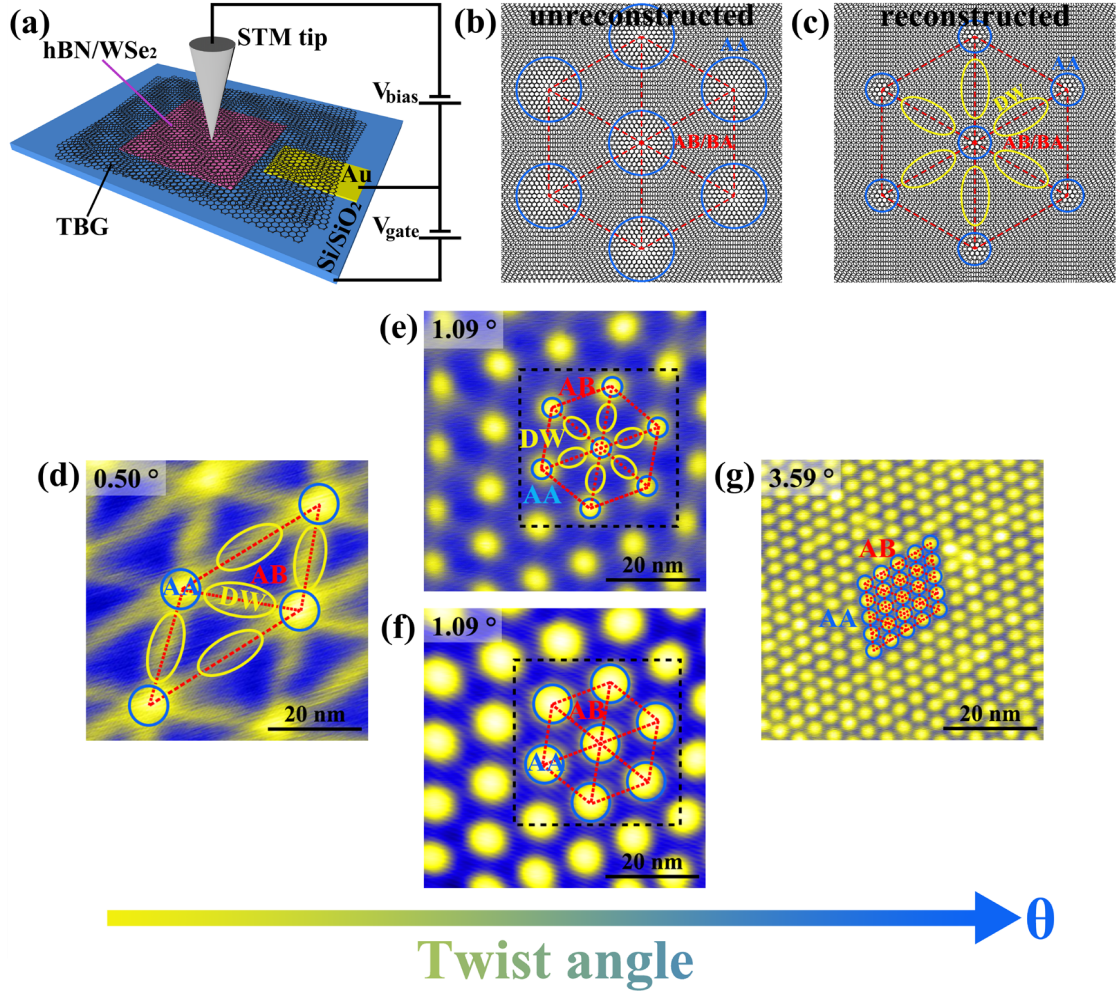

FIG. 1. (a) Schematic of the experimental device set-up. (b) and (c) Schematic of atomic registries in unreconstructed and reconstructed TBG, respectively. The blue and yellow circles represent the *AA* and *DW* regions of TBG, respectively. The rest of the red triangle are the *AB/BA* regions of TBG. (d)-(g) Representative STM images of the obtained TBG with different twist angles in our experiment. The scanning parameters are $V_{bias}$ = 0.6V, $I$ = 100pA (d), $V_{bias}$ = 0.5V, $I$ = 100pA (g), $V_{bias}$ = 0.3V, $I$ = 100pA (f), $V_{bias}$ = 0.8V, $I$ = 200pA (g).

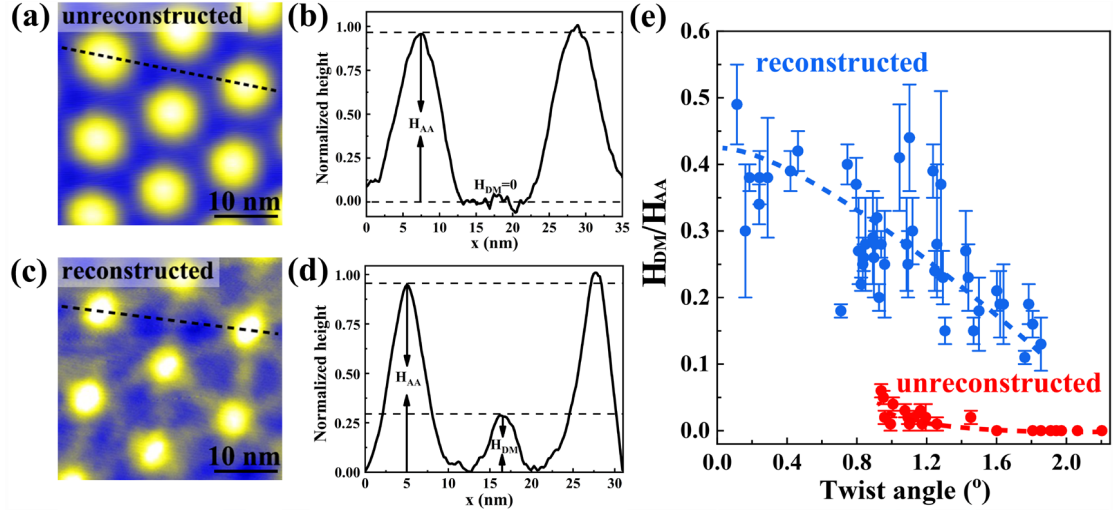

FIG. 2. (a) and (c) Zoom-in STM images of the TBGs in Fig. 1(f) and 1(e), respectively. (b) and (d) Height profiles along dashed lines in (a) and (c), respectively. The height of *AA* and *DW* regions are marked in the figures. (e) The measured $H_{DW}/H_{AA}$ as a function of the twist angles for different TBG devices. The red and blue points represent the results obtained in the unreconstructed and reconstructed TBG devices, respectively. The error bars reflect the standard error of the data.

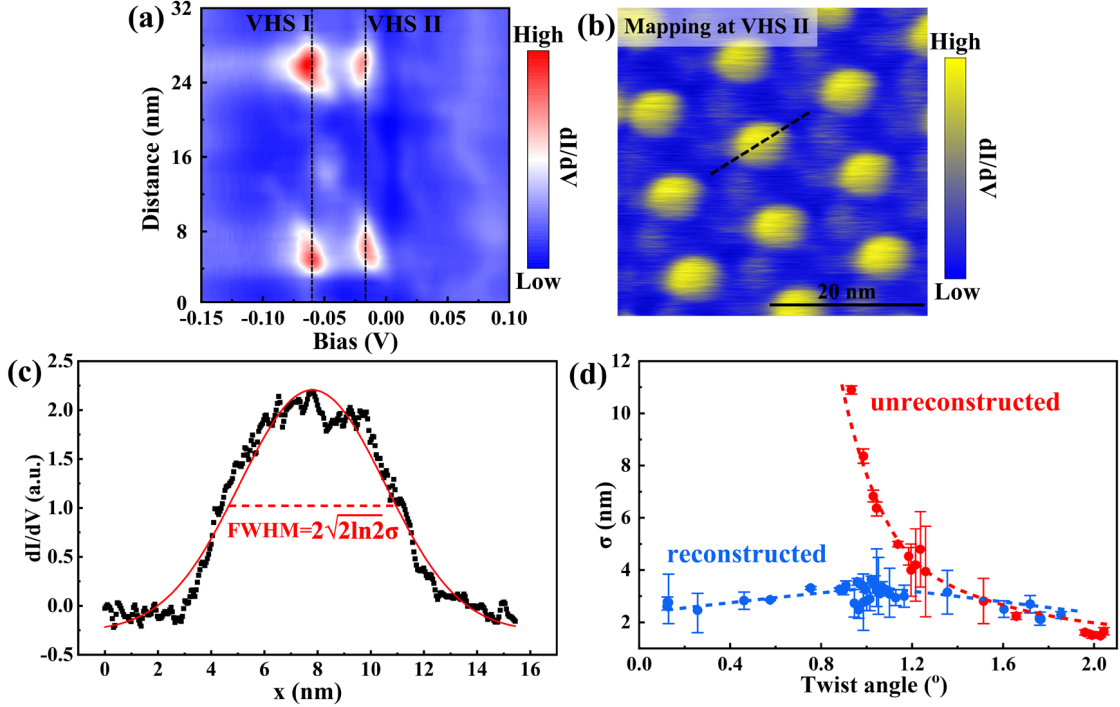

FIG. 3. (a) Representative STS spectra measured along the dashed line in Fig. 2(c). Two low-energy VHSs are observed in the *AA* regions of the TBG. (b) A typical STS map recorded at the energy of VHS II ($V_{bias}$ = -0.018V, $I$ = 40pA, $V_{gate}$ = 14V). (c) A profile line of the DOS at the energy of the VHS II along the dashed line in (b). The red solid curve represents a Gaussian fitting of the DOS distribution, and the red dashed line represents the full width at half maximum (FWHM) of the Gaussian fitting. (d) The averaged σ as a function of twist angles based on results obtained from part of the TBG devices in Fig. 2(e). The red and blue points represent results of the unreconstructed and reconstructed TBG devices, respectively. The error bars reflect the standard error of the data.

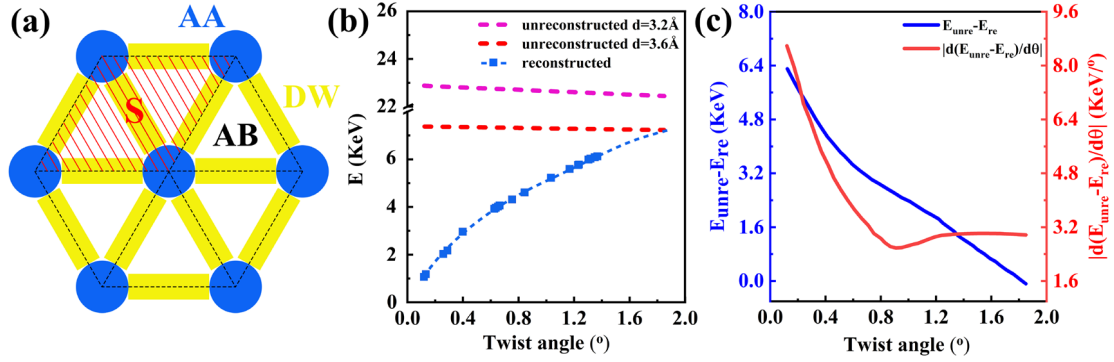

FIG. 4. (a) Schematic of stacking order assignments. The blue circles, yellow rectangles and white triangles represent *AA*, *DW* and *AB* regions, respectively. The red shade part represents the area of the moiré unit cell. (b) The calculated vdW stacking energy as a function of twist angles for both the reconstructed and unreconstructed TBG. The red and pink dashed curves represent the vdW stacking energy of the unreconstructed TBG with different interlayer distances. The blue squares represent the vdW stacking energy of the reconstructed TBG. (c) The difference of vdW stacking energy between the unreconstructed and reconstructed TBG (blue curve) and the absolute value of deviation of the vdW energy difference (red curve) as a function of twist angles.